\documentclass{article}
\usepackage{spconf,amsmath,graphicx}
\usepackage{setspace}
\usepackage{color}
\usepackage{fancyhdr}

\usepackage{amssymb}
\usepackage{amsmath}
\usepackage{bbm}
\usepackage{mathrsfs}
\usepackage{graphicx}
\usepackage{makecell}
 \usepackage{multirow}
 \usepackage{pdfpages}
\usepackage{subfigure}
\usepackage{tabularx}
\usepackage{booktabs}
\usepackage{textcomp,booktabs}
\usepackage{colortbl}

\usepackage{floatrow}
\newfloatcommand{capbtabbox}{table}[][\FBwidth]

\definecolor{mygray}{gray}{.9}

\title{TRANSAVS: END-TO-END AUDIO-VISUAL SEGMENTATION WITH TRANSFORMER}
 \name{Yuhang Ling$^{1}$ \ \  Yuxi Li$^{3}$ \ \  Zhenye Gan$^{3}$\ \  Jiangning Zhang$^{2,3}$ \ \  Mingmin Chi$^{1*}$ \ \   Yabiao Wang$^{2,3*}$}
  
  \address{$^{1}$ School of Computer Science, Fudan University \quad
       $^{2}$ Zhejiang University \quad
       $^{3}$ Youtu Lab, Tencent}
\begin{document}
\ninept
\maketitle

\begin{abstract}
    Audio-Visual Segmentation (AVS) is a challenging task, which aims to segment sounding objects in video frames by exploring audio signals. Generally AVS faces two key challenges: (1) Audio signals inherently exhibit a high degree of information density, as sounds produced by multiple objects are entangled within the same audio stream; (2) Objects of the same category tend to produce similar audio signals, making it difficult to distinguish between them and thus leading to unclear segmentation results. Toward this end, we propose TransAVS, the first Transformer-based end-to-end framework for AVS task. Specifically, TransAVS disentangles the audio stream as audio queries, which will interact with images and decode into segmentation masks with full transformer architectures. This scheme not only promotes comprehensive audio-image communication but also explicitly excavates instance cues encapsulated in the scene.    Meanwhile, to encourage these audio queries to capture distinctive sounding objects instead of degrading to be homogeneous, we devise two self-supervised loss functions at both query and mask levels, allowing the model to capture distinctive features within similar audio data and achieve more precise segmentation. Our experiments demonstrate that TransAVS achieves state-of-the-art results on the AVSBench dataset, highlighting its effectiveness in bridging the gap between audio and visual modalities.


\end{abstract}
\renewcommand{\thefootnote}{\fnsymbol{footnote}}
\footnotetext[1]{\ is the corresponding author.}

\vspace{-10pt}
\section{Introduction}
\vspace{-6pt}
\label{sec:intro}

Humans possess the remarkable ability to leverage audio and visual input signals to enhance the perception of the world \cite{percept}. For instance, we can identify the location of an object not only based on its visual appearance but also by the sounds it produces. This intrinsic connection has paved the way for numerous audio-visual tasks, including audio-visual correspondence \cite{avc1&ssl1, avc2&ssl2, avc3, avc6}, audio-visual event localization \cite{avel1, avel2, avel3, avel4, avel5}, audio-visual video parsing \cite{avvp1, avvp2, avvp3, avvp4}, and sound source localization \cite{avc1&ssl1, avc2&ssl2, ssl3}. However, the absence of pixel-wise annotations has limited these methods to frame or patch-level comprehension, ultimately restricting their training objectives to the classification of audible images.

Recently, a novel audio-visual segmentation (AVS) task was introduced in \cite{avs} with the aim of segmenting sounding objects corresponding to audio cues in video frames. This task is inherently a non-trivial one due to the two following challenges. Firstly, audio signals are information-dense, as they often contain sounds from multiple sources simultaneously. For example, in a concert, the sounds of instruments and human voices often become intertwined. This necessitates the disentanglement of audio signals at each timestamp into multiple latent components to effectively capture the unique sounding features of individual objects. Secondly, audio signals from objects of the same category often exhibit similar frequencies, such as in the case of the Husky and the Tibetan Mastiff. This ambiguity presents greater demands on the audio signal representation throughout the network to avoid inaccurately locating the sources of sound. However, the existing method \cite{avs} fails to address these challenges. 
Concrectly, it simply extracts audio features at each timestamp using an audio encoder, followed by the interaction with image embeddings through convolution, then generates the final prediction using an FPN-based scheme \cite{fpn} under the supervision of the standard segmentation~\cite{wu2023towards,li2023sfnet}.


\begin{figure}[!t]
    \includegraphics[width=\linewidth]{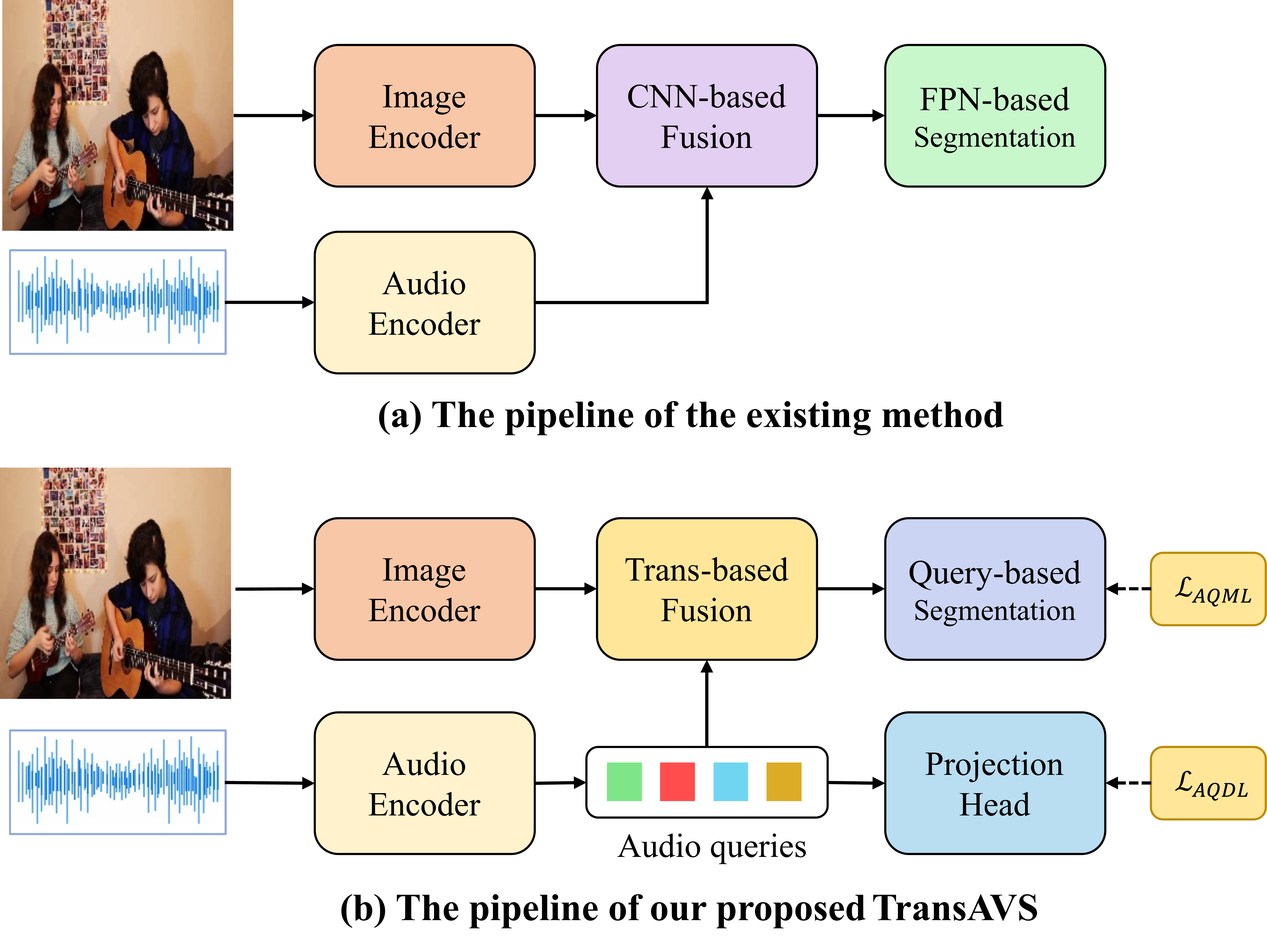}
    \caption{The pipeline comparison between: (a) the existing method and (b) our proposed TransAVS framework.}
    \label{fig:teasur}
    \vspace{-20pt}
\end{figure}

To this end, we propose a novel transformer-based end-to-end audio visual segmentation framework (\textbf{TransAVS}), drawing inspiration from the recent success of the transformer architecture in multi-modal learning \cite{trInmultimodal}. As depicted in Fig. \ref{fig:teasur}(b), TransAVS is a multi-modal transformer that leverages audio cues to guide both the fusion with visual features and segmentation. Concretely, first, to handle scenarios with multiple sounding objects, we disentangle the audio stream to initialize several audio queries, which encourages the model to explicitly attend to different objects, facilitating the acquisition of instance-level awareness and discrimination. Besides, we introduce two self-supervised loss functions at the query and mask levels, respectively. These functions play a pivotal role in optimizing the audio queries by encouraging heterogeneity during the learning process. This design empowers the model to discern and capture unique features embedded within similar audio streams, resulting in more precise segmentation.


\begin{figure*}[!t]
    \centering
    \includegraphics[width=\textwidth]{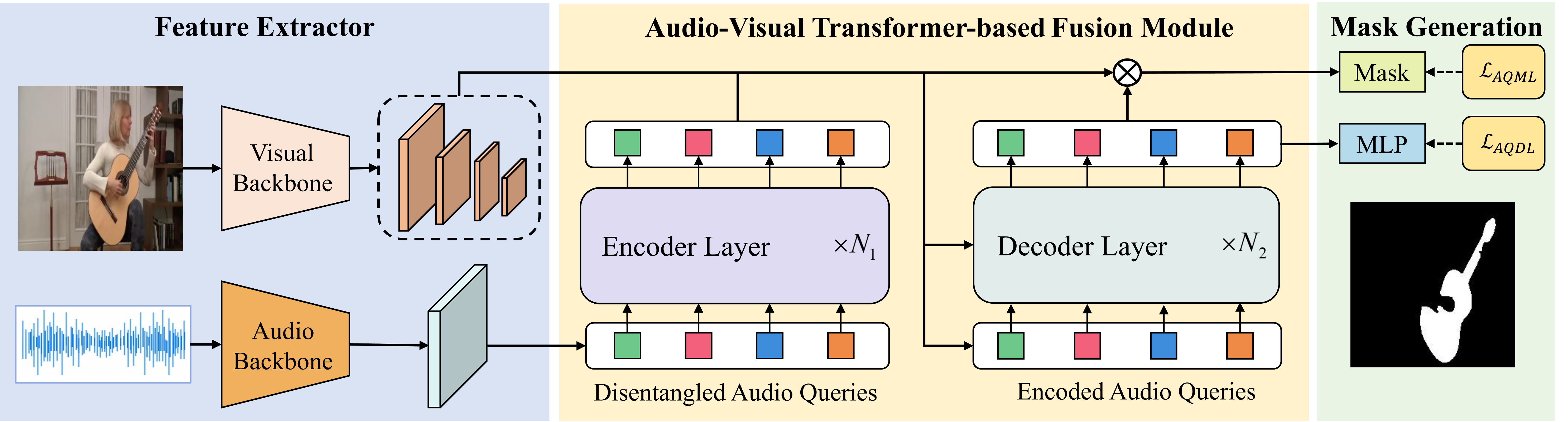}
    \caption{Architecture of the Transformer-based end-to-end framework, TransAVS. In this framework, the audio stream is disentangled into audio queries, guiding both the fusion with visual features and the segmentation process using the Transformer manner. To address the challenge of sound homogeneity among objects of the same category, we introduce two self-supervised loss functions at the query and mask levels. These innovative designs, distinct from existing method, not only enable the model to attain instance-level awareness and discrimination, but also distinguish and capture unique features embedded within similar audio streams, resulting in more precise segmentation.
    }
    \label{fig:main}
    \vspace{-18pt}
\end{figure*}

In summary, the main contributions of this paper are four-fold: (1) To the best of our knowledge, we are the first to introduce a multi-modal transformer-based framework to tackle the AVS task, leveraging the potent long-range modeling abilities to promote cross-modal interaction; (2) To guide the model towards perceiving and discriminating sounding objects at the instance level, we explicitly disentangle audio cues as audio queries. (3) To effectively address the issue of homogeneity of sounds among objects of the same category, we design two self-supervised loss functions, enabling the model to capture distinctive features within similar audio streams. (4) Qualitative and quantitative experimentation conclusively demonstrates the state-of-the-art performance of our method on the AVSBench dataset.

\vspace{-11pt}
\section{Methodology}
\label{sec:format}
\vspace{-8pt}

In this section, we will delve into the details of our proposed TransAVS framework. We begin by introducing the problem formulation in Section \ref{subsec:problemFormulation}, followed by a comprehensive explanation of the TransAVS architecture in Section \ref{subsec:Architecture}. Furthermore, we outline the design and rationale behind our self-supervised loss functions in Section \ref{subsec:sslf}. Lastly, Section \ref{subsec:infer} explains how TransAVS infers sounding object masks.

\vspace{-15pt}
\subsection{Problem Formulation}
\vspace{-5pt}
\label{subsec:problemFormulation}

\begin{table*}[t] \label{table:main result}
\centering
\caption{Quantitative comparison results of different methods on AVSBench. Our method outperforms the Baseline with a great gap in both the S4 and MS3 subsets across all visual backbones. Results of mean Jaccard index $\mathcal{M}_\mathcal{J}$ and F-score $\mathcal{M}_\mathcal{F}$ are reported.}

\resizebox{\textwidth}{!}{
\begin{tabular}{cccccccccccc}
\toprule
\multirow{2}{*}{Metric} & \multirow{2}{*}{Setting} & \multicolumn{2}{c}{SSL} & \multicolumn{2}{c}{VOS} & \multicolumn{2}{c}{SOD} & \multicolumn{2}{c}{Baseline\cite{avs}} & \multicolumn{2}{c}{Ours} \\ \cmidrule(r){3-4}\cmidrule(r){5-6}\cmidrule(r){7-8}\cmidrule(r){9-10}\cmidrule(r){11-12}
 &   & LVS\cite{sslIntro} & MSSL\cite{mssl} & 3DC\cite{3dc} & SST\cite{sst} & iGAN\cite{igan} & LGVT\cite{lgvt} & ResNet  & Pvt-v2 & ResNet & Swin-base    
 \\ \midrule
\multirow{2}{*}{$\mathcal{M}_\mathcal{J}$} & Single-source(S4) & 37.9 & 44.9 & 57.1 & 66.3 & 61.6 & 74.9   & 72.8 & 78.7  & 83.1 &  \textbf{89.4}             
\\
& Multi-source(MS3) & 29.5 & 26.1 & 36.9 & 42.6 & 42.9 & 40.7 & 47.9 & 54.0 & 58.9 & \textbf{63.5}            
\\ \midrule
\multirow{2}{*}{$\mathcal{M}_\mathcal{F}$} & Single-source(S4)
& 51.0 & 66.3 & 75.9 & 80.1 & 77.8 & 87.3 & 84.8 & 87.9 & 90.6 & \textbf{94.2}            
\\
& Multi-source(MS3) & 31.0 & 36.3 & 50.3 & 57.2 & 54.4 & 59.3 & 59.3 & 64.5 & 72.9 & \textbf{75.2}     
\\
\bottomrule
\end{tabular}
}
\vspace{-15pt}
\label{table:MainResult}
\end{table*}

For the AVS task, the input data comprises a sequence of video frames $\mathcal{V}=\{v_i \}_{i=1}^T$, where $v_i$ $\in\mathbb{R}^{3\times H_v \times W_v}$, and $T$-second audio stream $\mathcal{A}$. The goal of AVS is to segment all sounding objects in each frame $v_i$ under the acoustic guidance $\mathcal{A}$. The segmentation results are binary masks $\mathcal{M}=\{m_i\}_{i=1}^T$, where $m_i\in \{0,1\}^{H_v \times W_v}$, with `1' indicates  sounding objects while `0' corresponds to background or silent objects.
\vspace{-15pt}
\subsection{The Architecture of AUST}
\vspace{-6pt}
\label{subsec:Architecture}
Our TransAVS framework consists of 3 modules: (1) a feature extractor which is responsible for extracting multi-scale image features $F_v$ and audio features $F_a$; (2) an audio-visual transformer-based fusion module, which disentangles $F_a$ into audio queries $A_q$ and fuses with $F_v$ in a transformer fashion; (3) a mask generation module predicting binary masks with probability of sounding objects.
\vspace{-13pt}
\subsubsection{Feature Extractor}
\vspace{-6pt}
\noindent\textbf{Visual Feature:} Taking one frame $v_i$ in $\mathcal{V}$ as input, a pretrained visual backbone is employed to extract visual features. To exploit semantic information in different levels, we extract vision features in 3 scales $F_v=\{f_{v}^i\}_{i=1}^3$, where $f_{v}^i \in \mathbb{R}^{C_v \times \frac{H_v}{2^{4-i}} \times \frac{W_v}{2^{4-i}}}$, $C_v$ is the channel dimension depending on different encoders. We also upsample $f_{v}^3$ as $f_v^4 \in \mathbb{R}^{C_v \times H_v \times W_v}$ for later mask generation. 

\noindent\textbf{Audio Feature:} Given the audio clip $\mathcal{A}$, we first process it to a spectrogram via the short-time Fourier transform, then pass it to a pretrained audio backbone VGGish \cite{vggish} to obtain the audio embedding $F_a \in \mathbb{R}^{T\times d}$.
\vspace{-13pt}
\subsubsection{Audio-visual Transformer-based Fusion Module}
\vspace{-4pt}
As previously mentioned, we first disentangle audio features $F_a$ into audio queries $A_q$ to facilitate the model's learning of instance-level awareness and discrimination, then adopt attention mechanism to establish long-range connection between audio cues and visual features. Technically, we begin by projecting $F_a$ into $N$ independent queries with a linear transform $W_1 \in \mathbb{R}^{N\times T}$:
\setlength{\abovedisplayskip}{4pt}
\setlength{\belowdisplayskip}{4pt}
\begin{align}
    A_q^0 =W_1 F_{a}
\end{align}
then input them into $N_1$ encoder layers to capture their dependency. Concretely, at the $n$-th layer: 
\setlength{\abovedisplayskip}{2pt}
\setlength{\belowdisplayskip}{3pt}
\begin{align}
    \text{Atten}(Q,K,V) &= \text{softmax}(\frac{QK^T}{\sqrt{d}})V \\
    A_q^{n+1} = \text{Atten}(A_q^n &W_Q^n, A_q^n W_K^n , A_q^n W_V^n ) + A_q^n     
    \vspace{-14pt}
\end{align}
where $A_q^n \in \mathbb{R}^{N\times d}$, $W_{Q}^{n}$, $W_{K}^{n}$ and $W_{V}^{n}$ respectively, represent the Q, K, V transform matrix at the $n$-th layer in $\mathbb{R}^{d\times d}$, following the standard attention scheme \cite{vanila-transformer}. 

\vspace{-2pt}
After $N_1$ encoder layers, the output $A_q^{N_1}$ conveys different audio components information, providing the network with guidance for attending to different sounding regions during the cross-modal fusion process within following $N_2$ decoder layers. Specifically, at the $l$-th layer, audio queries $A_q=A_q^{N_1}$ serves as query while image features $f_1$, $f_2$, $f_3$ act as keys and values in a round-robin fashion:
\setlength{\abovedisplayskip}{3pt}
\setlength{\belowdisplayskip}{3pt}
\begin{align}
    i &=\  (l\ \text{mod}\ 3) +1\\
    F_{av}^l &=
\begin{cases}
\ A_q\ & \text{if $l=1$ } \\
\ F_{av}^l & \text{in other cases }
\end{cases} \\
    F^{l+1}_{av} = \text{Atten}&(F^l_{av} W_Q^{l}, f_{v}^i W_K^{l} , f_{v}^i W_V^{l} )+ F^l_{av}
\end{align}
where `mod' denotes modulo operation, $F^{l}_{av} \in \mathbb{R}^{N\times d}$, $W_{Q}^{l}$, $W_{K}^{l}$ and $W_{V}^{l}$ denote the Q, K, V transformation matrix at the $l$-th layer, respectively. This approach not only establishes long-range connections between the audio stream and visual frames but also compels the model, through audio queries, to be aware of and discriminate sounding objects at the instance level.

\vspace{-11pt}
\subsubsection{Mask Generation}
\vspace{-4pt}
Based on the fused feature $F_{av}=F_{av}^{N_2}$ and the image embedding $f_v^4$, the mask generation module predicts segmentation masks $\mathcal{M}=\{M_i\}_{i=1}^N$ with the probability $\mathcal{P}=\{p_i\}_{i=1}^N$ of sounding objects. 

Technically, for binary mask $\mathcal{M}$ generation, we apply 1 $\times$ 1 convolution denoted as $W_2$ on $f_v^4$ to adjust the channel dimension to $d$, then multiply it with $F_{av}$ followed by a sigmoid function:
\setlength{\abovedisplayskip}{5pt}
\setlength{\belowdisplayskip}{5pt}
\begin{align}
    \mathcal{M} &= \text{sigmoid}(F_{av}W_2f_v^4)
\end{align}

Meanwhile, to calculate the probability $\mathcal{P}$, a classifier $g \in \mathbb{R}^{d \times K}$ and softmax function are utilised:
\setlength{\abovedisplayskip}{5pt}
\setlength{\belowdisplayskip}{5pt}
\begin{align}
    \mathcal{P} &= \text{softmax}[g(F_{av})]
\end{align}
where $K=2$ is the number of category. $\mathcal{M}$ and $\mathcal{P}$ are paired with audio queries $A_q$ as $\mathcal{Z}={\{z_i=(q_i, m_i,p_i)\}}_{i=1}^N$ for optimization and inference.

\vspace{-12pt}
\subsection{Self-Supervised Loss Functions}
\vspace{-4pt}
\label{subsec:sslf}

As mentioned before, objects of the same category tend to produce similar sound frequencies, resulting in a significant degree of homogeneity that can hinder the model's performance. Toward this, we propose two loss functions at query and mask level, namely the Audio Query Distance Loss (\textbf{AQDL}) and the Audio Query Mask Loss (\textbf{AQML}), with the goal of increasing heterogeneity and thus enhancing segmentation accuracy.


To be specific, AQDL, denoted as $\mathcal{L}_{AQDL}$, is a penalty on $q_i$ that predicts sounding objects but getting too close to each other, indicating that they have a high similarity with less clear guidance:
\vspace{-16pt}
\begin{figure*}[t]
    \includegraphics[width=\textwidth
    ]{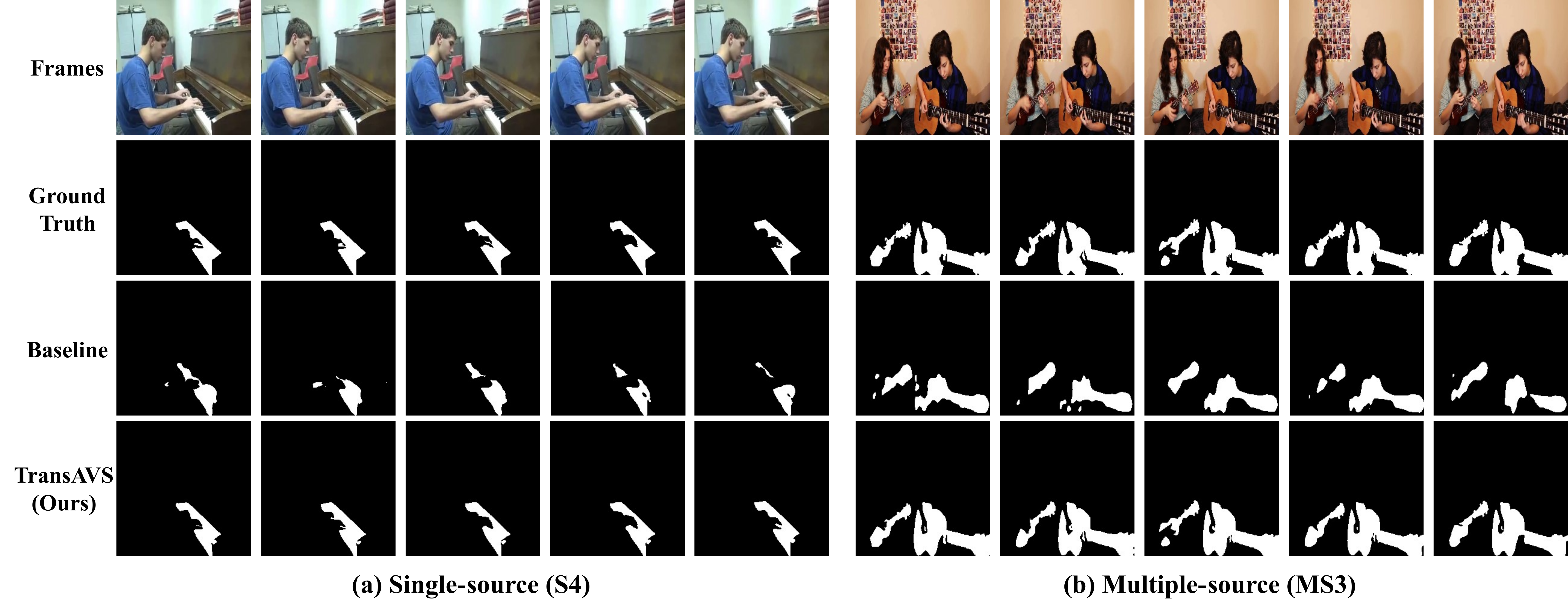}
    \caption{Qualitative comparison between the Baseline and our proposed TransAVS. Our method's instance-level awareness and discrimination, enabling it to distinguish between individual sounding sources, significantly contribute to the more precise segmentation, which is evident in the subfig (b), where TransAVS effectively delineates the shape of the sounding source (guitars) while discarding the silent objects (hands).}
    \label{fig:quali}
    \vspace{-24pt}
\end{figure*}
\setlength{\abovedisplayskip}{10pt}
\setlength{\belowdisplayskip}{3pt}
\begin{align}
    \mathcal{L}_{AQDL}&=\frac{2}{n_1(n_1+1)}\sum_{i=1}^{n_1}\sum_{j=i+1}^{n_1} d(q_i,q_j)\\
    d(q_i,q_j)&=
    \begin{cases}
\frac{1}{||h(q_i)-h(q_j)||_2^2}\ & \text{if } ||h(q_i)-h(q_j)||_2^2<d_0 \\
\ \ \ \ \ \ 0\ & \text{in other cases }
\end{cases}
\end{align}
where $h$ represents a projection head in $\mathbb{R}^{d\times d}$, $q_i$ and $q_j$ are elements of the set $S_{1}=\{z_k|p_k>\delta_{1}\}$, $\delta_{1}$ is the confidence threshold of $\mathcal{L}_{AQDL}$, $n_1$ is the cardinality of  $S_{1}$, $||$·$||_2$ is the $\mathcal{L}_2$ norm, and $d_0$ is the threshold for $d(a_i,a_j)$. AQDL promotes heterogeneity by restricting queries from getting too close.

On the other hand, AQML encourages $q_i$ to predict exclusive sounding masks $m_i$ as much as possible. It focuses on the intersecting pixels between different sounding-object masks:
\setlength{\abovedisplayskip}{3pt}
\setlength{\belowdisplayskip}{3pt}
\begin{align}
    \mathcal{L}_{AQML}&=\frac{2}{n_2(n_2+1)}\sum_{i=1}^{n_2}\sum_{j=i+1}^{n_2} I(m_i,m_j)   \\     
    I(m_i,m_j)&=\frac{1}{2HW}[\text{Bin}(m_i)\odot m_j + m_i\odot \text{Bin}(m_j)]
    \label{eq:inter}
\end{align}
where $m_i$ and $m_j$ belong to the set $S_{2}=\{z_k|$ $p_k>\delta_{2}\}$, $\delta_{2}$ is the confidence threshold of $\mathcal{L}_{AQML}$, $n_2$ is the cardinality of $S_{2}$, `Bin' denotes the binary operation with threshold set $0.5$ and $\odot$ is the Hadamard product. AQML forces queries to attend to different parts of images, thus reducing their heterogeneity.

Besides AQDL and AQML, we also use 2 supervised segmentation losses, including focal classification loss $\mathcal{L}_{class}$\cite{focal} and dice loss $\mathcal{L}_{dice}$\cite{dice}. All losses are linearly combined as optimization objective during our end-to-end training process: 
\setlength{\abovedisplayskip}{3pt}
\setlength{\belowdisplayskip}{3pt}
\begin{align}\label{eq:total loss}
\mathcal{L}=\lambda_{1}\mathcal{L}_{AQDL}+\lambda_{2}\mathcal{L}_{AQML}+\lambda_{3}\mathcal{L}_{class}+\lambda_{4}\mathcal{L}_{dice}
\end{align}
\vspace{-26pt}
\subsection{Inference Stage} \label{subsec:infer}
\vspace{-4pt}
During the inference, TransAVS predicts each pixel at location $(x,y)$ based on $\mathcal{Z}$:
\setlength{\abovedisplayskip}{2pt}
\setlength{\belowdisplayskip}{2pt}
\begin{align}
    \mathop{\arg\max}_{c_i}\  p_i(c_i)\times m_i(x,y) \\
    c_i=\mathop{\arg\max}_{c \in \{{1,...,K\}}}\  p_i(c)\ \ \  \forall z_i \in z
\end{align}
that is, when and only when both the class probability $p_i(c_i)$ and the mask prediction probability $m_i(x,y)$ are high enough will a pixel $(x, y)$ be assigned to $z_i$ .

\vspace{-6pt}
\section{Experiments}
\vspace{-2pt}
\subsection{AVSBench Dataset}\label{sec:dataset}
\vspace{-6pt}
All videos in AVSBench dataset are trimmed into 5 seconds and separated into 2 subsets based on the number of sound source: \textit{single-source sound segmentation (S4)} and \textit{multiple-sound source segmentation (MS3)}. Then each video is divided into 5 non-overlapping 1-second clips, each clip is sampled one frame. As shown in Table \ref{table:dataset}, only the first sampled frame in the training split of S4 is annotated while all frames in other split are annotated. S4 contains 23 classes (Cls.), covering  sounds from humans, animals, vehicles, and musical instruments. Each video in the MS3 subset includes two or more categories from the S4 subset.
\vspace{-16pt}
\begin{table}[h]
\centering
\normalsize
\caption{AVSBench dataset statistics. For S4 training split, one annotation per video while all others contain 5 annotations per video.}
\vspace{-10pt}
\resizebox{\linewidth}{!}{
\begin{tabular}{ccccc}
\toprule
Subset           & Cls. & Videos & Train/Valid/Test & Annotated frames \\ \midrule
Single-source(S4)    & 23            & 4932         & 3,452\ /\ 740\ /\ 740    & 3452\ /\ 3700\ /\ 3700         \\ 
Multi-source(MS3) & 23            & 424          & 296\ /\ 64\ /\ 64        & 1480\ /\ 320\ /\ 320          \\ \bottomrule
\end{tabular}
}
\vspace{-23pt}
\label{table:dataset}
\end{table}
\subsection{Implementation Details}
\vspace{-4pt}
For the visual backbone, we choose 2 representative ones: the standard CNN-based ResNet\cite{resnet} backbone R101 and Transformer-based Swin-Transformer\cite{swintrans} backbone swin-base. R101 is pretrained on ImageNet-1K\cite{imagenet1k} while swin-base on ImageNet-22K. For the loss weights, we set $\lambda_1=\lambda_2=2.0$ and $\lambda_3=\lambda_4=5.0$. For the optimizer, we use AdamW \cite{adamw} with an initial learning rate of 0.0001 for both R101 and swin-base backbones. A learning rate multiplier of $0.1$ is also applied. All models are trained with 8 3090 GPUs for 90k iterations with a batch size of $8$.
\vspace{-12pt}
\subsection{Main Results}
\vspace{-4pt}
Since AVS is a newly proposed problem, we compare our network with baseline in \cite{avs} and methods from three related tasks, including sound source localization (SSL), video object segmentation (VOS), and salient object detection (SOD). For each task, we report the results of two SOTA methods on AVSBench dataset, \textit{i.e.}, LVS\cite{sslIntro} and MSSL\cite{mssl} for SSL, 3DC\cite{3dc} and SST\cite{sst} for VOS, iGAN\cite{igan} and LGVT\cite{lgvt} for SOD. To ensure fairness, all backbones of these methods were pretrained on the ImageNet-1K\cite{imagenet1k} dataset.

\noindent\textbf{Quantitative Comparison.} Given a test frame, we denote the predicted mask as $m$ and ground truth as $y$, the Jaccard index $\mathcal{J}$ \cite{jaccard} and F-score $\mathcal{F}$ are used to measure region similarity and contour accuracy, respectively:
\setlength{\abovedisplayskip}{3pt}
\setlength{\belowdisplayskip}{3pt}
\begin{align}
    \text{\ \ \ \ \ \ }\mathcal{J} &= \frac{m \cap y }{m \cup y}  \\
    \mathcal{F} &= \frac{(1+\beta^2) \times \text{precision} \times \text{recall} }{\beta^2 \times \text{precision} + \text{recall}}
\end{align}
where $\beta=0.3$. We use $\mathcal{M}_\mathcal{J}$ and $\mathcal{M}_\mathcal{F}$ to denote the mean metric values over the whole test dataset. The quantitative results is shown in Table \ref{table:MainResult}. It is evident that our proposed approach consistently outperforms existing methods in both subsets across all visual backbones. Even in the S4 subset, where the Baseline achieves high values on the $\mathcal{M}_\mathcal{J}$ metric (72.8 with ResNet50 and 78.7 with Pvt-v2), our proposed TransAVS still shows improvements: 10.3 points higher with ResNet and 10.7 points higher with Pvt-v2. We attribute this improvement to our transformer framework which uses audio queries to explicitly learn instance-level awareness and discrimination of sounding objects, as well as our loss functions that increase heterogeneity among sounds from objects of the same category. These design choices allow our model to exploit important audio cues to gain better segmentation.   

\noindent\textbf{Qualitative Comparison.} We provide some qualitative examples from both S4 and MS3 in Fig. \ref{fig:quali}. The segmentation results clearly demonstrate that our method outperforms the baseline. We believe that our method's instance-level awareness and discrimination, enabling it to distinguish between individual sounding sources, significantly contribute to the more precise segmentation. This is especially evident in the Fig. \ref{fig:quali}(b), where TransAVS effectively delineates the shape of the sounding source (guitars) while discarding the silent objects (hands).
\vspace{-10pt}
\subsection{Ablation Study}
\vspace{-4pt}
\begin{table}[]
\centering
\caption{Ablation study in both S4 and MS3 setting validates that our key designs are essential for the performance of TransAVS.}
\resizebox{\linewidth}{!}{
\begin{tabular}{c|c|cc|cc}
\toprule
\multirow{2}{*}{$N$} & \multirow{2}{*}{Mode of $\delta_1$ and $\delta_2$} & \multicolumn{2}{c|}{S4} & \multicolumn{2}{c}{MS3} \\ \cline{3-4} \cline{5-6}
  &    & $\mathcal{M}_\mathcal{J}$  &$\mathcal{M}_\mathcal{F}$    & $\mathcal{M}_\mathcal{J}$  & $\mathcal{M}_\mathcal{F}$  \\ 
  \hline \rule{0pt}{8pt}
100    & increasing $\delta_1$ and $\delta_2$   & 80.8    & 87.4    & 56.2    & 70.3  \\
500      & increasing $\delta_1$ and $\delta_2$   & 81.2       & 89.4       & 57.1       & 71.8       \\
 \hline \rule{0pt}{8pt}
300   & only fixed $\delta_1=0.6$  & 80.4  & 87.9   & 56.3       & 70.1       \\ 
300   & only fixed $\delta_2=0.6$ & 80.5  & 87.7       & 56.2  & 70.4    \\
 \hline
\rowcolor{mygray} 300 & increasing $\delta_1$ and $\delta_2$ & \textbf{83.1}  & \textbf{90.6}  & \textbf{58.9} & \textbf{72.9} \\ \bottomrule
\end{tabular}
}
\label{table:ab}
\end{table}

In Table \ref{table:ab}, we verify the effectiveness of each key design in the proposed method with ResNet backbone on both S4 and MS3 subsets. Based on the first 2 rows and the last row, our results show that the optimal performance obtained when the number of queries $N$ was set to 300. In the third and fourth rows, $\delta_1$ and $\delta_2$ are set to $0.6$, respectively. Both fixed mode show a notable decrease compared with the increasing mode adapted in the last row:
\setlength{\abovedisplayskip}{4pt}
\setlength{\belowdisplayskip}{4pt}
\begin{align}
\delta_{i}&=a+\frac{b-a}{18}\times \lfloor\frac{\text{iterations}}{n_{iter}}\rfloor, i \in \{{1,2}\}
\end{align}
where $a=0.55$, $b=0.65$, $n_{iter}=5000$. We hypothesize that this phenomenon can be explained as follows: as TransAVS becomes increasingly confident about its mask prediction, a fixed threshold may not penalize audio queries with less confidence at earlier stages while pushing too many audio queries away at later epochs. This leads to poorer performance compared to using an increasing threshold.


\section{Conclusion}

In this paper, we introduce TransAVS, the first transformer-based framework for the AVS task. We disentangle audio as audio queries to explicitly guide the model in learning instance-level awareness and discrimination of sounding objects. Additionally, we design self-supervised loss functions to address the homogeneity of sounds within the same category. Experimental results on the AVSBench dataset demonstrate that TransAVS achieves SOTA performance in both the S4 and MS3 subsets, demonstrating the effectiveness of TransAVS in bridging the gap between audio and vision modalities.

\vfill\pagebreak

\clearpage
\bibliographystyle{IEEEbib}
\bibliography{strings,refs}

\end{document}